\documentstyle[12pt,psfig]{article}   
\newcommand {\gsim}{\mathrel{\hbox{\rlap{\lower.55ex \hbox {$\sim$}}            
            \kern-.3em \raise.4ex \hbox{$>$}}}}
\newcommand {\lsim}{\mathrel{\hbox{\rlap{\lower.55ex \hbox {$\sim$}}            
            \kern-.3em \raise.4ex \hbox{$<$}}}}
\newcommand {\vectwo}[2] {\left(\begin{array}{c}#1\\#2\end{array}\right)}

\newcommand {\beq} {\begin{equation}}
\newcommand {\eeq} {\end{equation}}

\newcommand {\ncl} {non-contractible loop}

\newcommand {\sm} {standard model}
\newcommand {\ew} {electroweak}
\newcommand {\YMth} {Yang-Mills theory}

\newcommand {\YMHth} {Yang-Mills-Higgs theory}

\newcommand {\Istar} {I$^{\star }$}

\newcommand {\Sstar} {S$^{\star }$}

\newcommand {\D}{\partial}

\def\R  {{\rm I\kern-.15em R}}
\def\L  {{\rm I\kern-.25em L}}
\def\u  {{\bf u}}
\def\v  {{\bf v}}
\def\w  {{\bf w}}
\def\id {{\rm 1\kern-.12em
\rule{0.3pt}{1.5ex}\raisebox{0.0ex}{\rule{0.1em}{0.3pt}}}}
\newcounter{saveeqn}
\newcommand{\alpheqn}{\addtocounter{equation}{1}\setcounter{saveeqn}{\value{equation}}%
\setcounter{equation}{0}%
\renewcommand{\theequation}{%
             \mbox{\arabic{saveeqn}\alph{equation}}}}%
\newcommand{\reseteqn}{\setcounter{equation}{\value{saveeqn}}%
\renewcommand{\theequation}{\arabic{equation}}}
\newlength{\capowidth}
\newlength{\length}

\newcommand{\casecenter}[1]{
    \setlength{\capowidth}{12.cm}
  \begin{center}
    \parbox{\capowidth}{
    \settowidth{\length}{#1}
    \ifdim\capowidth>\length
      \begin{center} #1 \end{center}
    \else
       #1
    \fi}
\end{center}}
\renewcommand{\caption}[1]{
  \refstepcounter{figure}
  \casecenter{\footnotesize \sloppy
        {\bf Fig. \arabic{figure}.} #1}}

\begin{document}

\begin{titlepage}
\hspace*{\fill} hep-ph/9606481  
\newline
\hspace*{\fill} KA--TP--13--1996
\newline
\hspace*{\fill} June 1996
\begin{center}
\vspace{3\baselineskip}
{\Large \bf Construction of a new constrained instanton \\in \YMHth}\\
\vspace{1\baselineskip}
{\large F. R. Klinkhamer and J. Weller } \\
\vspace{1\baselineskip}
 Institut f\"ur Theoretische Physik\\ Universit\"at Karlsruhe\\
 D--76128 Karlsruhe\\ 
\vspace{3\baselineskip}
{\bf Abstract} \\
\end{center}
{\small
\noindent A self-consistent ansatz is presented for a
four-dimensional euclidean solution
(instanton) in the vacuum sector of constrained $SU(2)$ \YMHth.
}
\vspace{3\baselineskip}
\begin{tabbing}
PACS numbers \= : \= 11.15.Kc, 12.15.-y, 11.15.Bt, 12.38.Cy                \\
keywords     \> : \> classical solution, instanton, anomaly, summability \\
\end{tabbing}
\vfill
published in Nucl. Phys. {\bf B 481} (1996), 403
\end{titlepage}

\section{Introduction}
A topological argument has been given \cite{K93a} for the existence of a new
instanton\footnote[1]{Here, and in the following,
the term ``instanton'' refers to {\em any\/} localized,
finite action solution of the general field equations belonging to the
euclidean action of the theory considered, 
not exclusively to solutions of certain first order self-duality
conditions \cite{BPST75,ADHM78}.} in the va\-cu\-um sector of euclidean $SU(2)$
Yang-Mills-Higgs (YMH) theory, with a constraint added to fix the scale of the
solution. The existence argument was based on the construction of a suitable
\ncl ~(NCL) of 4-dimensional field configurations. The basic idea now is that
by making appropriate modifications of the configuration at the top of this
particular NCL one arrives at an exact solution of the field equations. The
present paper gives the resulting self-consistent ansatz for this
new constrained instanton \Istar. Technically the construction of \Istar ~is  more
complicated than for the related 3-dimensional sphaleron solution
\Sstar ~\cite{K93b}, but the method is essentially the same. 

Most likely, \Istar ~is the lowest action constrained instanton solution in
the vacuum sector of euclidean
$SU(2)$ YMH theory. As such it is expected to be of fundamental importance for
the quantum field theory of the electroweak interactions. In particular, \Istar ~is
believed to play a role in the asymptotics of Feynman perturbation theory
\cite{L77}. 

The outline of this paper is as follows. Section 2 describes briefly the theory
considered. Section 3 presents the ansatz and Section 4 lists the
symmetries, which are crucial for the derivation. Section 5 gives the
action evaluated for the ansatz fields and Section 6 explains how
the ansatz leads to having a non-trivial solution of the field
equations. Section 7, finally, gives a brief discussion of the potential
physics applications mentioned above. There are also three
appendices. Appendix A puts the NCL of
\cite{K93a} in a form appropriate for the construction of the ansatz of Section 3. 
Appendix B gives the transformations of the ansatz functions under the
residual gauge symmetry discussed in Section 4. 
Appendix C, finally, gives the ansatz actiondensity of Section 5 in a compact notation.

\section{Theory}
Consider a classical  \YMth, with non-Abelian gauge group $SU(2)$ and
gauge coupling constant $g$, coupled to a Higgs scalar field in the fundamental
representation, with vacuum expectation value $v$ and quartic coupling constant
$\lambda$.  The two mass scales of the theory are 
$M_{\rm W} \equiv \frac{1}{2}g\,v$ for the three W bosons and $M_{\rm H} \equiv
\sqrt{2\lambda}\,v$ for the single Higgs scalar. 
The total euclidean action is  
\beq
          A = A_{\mbox {\rm \tiny YMH}} + A_{\mbox {\rm \tiny C}}\, ,
\label{eq:A}
\eeq
with the Yang-Mills-Higgs action
\beq
     A_{\mbox {\rm \tiny YMH}} = \int\limits_{\R^4} d^4x
          \left[\,- \, \frac{1}{2 \, g^2}\, {\rm Tr} \: W_{\mu \nu}^2
                 + |D_\mu \Phi|^2
                 + \lambda \left( |\Phi|^2 - \frac{v^2}{2} \right)^2 \; \right],
\label{eq:AYMH}
\eeq
where $W_{\mu\nu} \equiv \partial_\mu W_\nu -\partial_\nu W_\mu +
[W_\mu,W_\nu]$, $D_\mu \Phi \equiv (\partial_\mu + W_\mu)\Phi$, $W_\mu \equiv
W_\mu^a \sigma^a/(2i)$ and $\sigma^a$ are the standard Pauli matrices, and a constraint
term \cite{A81}
\beq 
     A_{\mbox {\rm \tiny C}} = \frac{\kappa}{g^2} \left( \frac{1}{M_W^4}
                 \int\limits_{\R^4}d^4x \: O_8
                 - \frac{8\pi^2c}{(M_{\rm W}\varrho)^4} \right),
\label{eq:ac}
\eeq 
with $\kappa$ an arbitrary positive constant (Lagrange multiplier) and
$\varrho$ the ``size'' of the field configuration.
Specifically, we choose for the constraint operator
\beq
       O_8 = q_{\rm P}^2\; ,
\label{eq:o8}
\eeq
where $q_{\rm P}$ is the Pontryagin density
\beq
 q_{\rm P} \equiv - \,\frac{1}{4}\,\epsilon_{\kappa \lambda \mu \nu}\; {\rm
                          Tr}\: W_{\kappa\lambda}W_{\mu\nu}\; ,
\eeq
and for the numerical constant $c$ we take the value
\beq
       c = \frac{288}{21} \; .
\label{eq:c}
\eeq
This
particular choice for $c$ reproduces the usual scale parameter of a BPST
\cite{BPST75} instanton--anti-instanton pair at infinite separation.
Of course, there are many other constraint operators $O_d$ possible, as long as
they have canonical mass dimension $d>4$.
Also, there might, in principle, exist solutions
for which the single constraint operator
(\ref{eq:o8}) does not suffice and further terms need to be added.  

Having defined the theory we proceed in three steps. First, we obtain for a given
positive value of $\kappa$ a solution $W^\star,\Phi^\star$ of the field
equations resulting from variations $\delta W$, $\delta \Phi$ of the total
action $A$. Second, we solve for the scale $\varrho^\star =
\varrho^\star(\kappa,W^\star,\Phi^\star)$, so that $A_{\mbox {\rm \tiny C}}^\star = 0$ and
$A^\star = A_{\mbox {\rm \tiny YMH}}^\star(\kappa,W^\star,\Phi^\star)$, and
eliminate $\kappa$ between $\varrho^\star$ and $A_{\mbox {\rm \tiny YMH}}^\star$. Third, with these values $\varrho^\star$ and $A_{\mbox {\rm \tiny 
YMH}}^\star(\varrho^\star,W^\star,\Phi^\star)$, we
integrate over the collective coordinate $\varrho$ in the path integral of the particular Green's function
considered. Further details may be found in \cite{K93a,A81}.

Here, we focus on the first crucial step, namely to discover a non-trivial
solution of the field equations belonging to the euclidean action
(\ref{eq:A}). More specificially, we look for a finite action solution 
with Pontryagin index (topological charge)
\beq
        Q_{\rm P} \equiv \frac{1}{8\pi^2} \: \int\limits_{\R^4}d^4x \: q_{\rm P}
\label{eq:Qp}
\eeq
vanishing, i.~e. a new constrained instanton in the vacuum sector.

\section{Ansatz}
Define cylindrical coordinates $\rho$, $\varphi$, $z$ and $\tau$ in
terms of the cartesian coordinates
\[
            (x_1,x_2,x_3,x_4) \equiv (\rho
                                     \cos{\varphi},\rho\sin{\varphi},z,\tau)
\]
and introduce a triad of matrices
\begin{eqnarray*}
                \u &\equiv&  \sin{\varphi}\; \tau^2 + \cos{\varphi}\;\tau^1\\
                \v &\equiv&  \cos{\varphi}\; \tau^2 - \sin{\varphi}\;\tau^1\\
                \w &\equiv& \tau^3,
\end{eqnarray*}
with $\tau^a \equiv \sigma^a/(2i)$ and $\sigma^a$ the Pauli matrices
\[
    \sigma^1 \equiv \left( \begin{array}{cc} 0 & 1 \\ 1 & 0 \end{array} \right)
    \qquad
    \sigma^2 \equiv \left( \begin{array}{cc} 0 & -i \\ i & 0 \end{array} \right)
    \qquad
    \sigma^3 \equiv \left( \begin{array}{cc} 1 & 0 \\ 0 & -1 \end{array} \right).
\]
The ansatz is then given by 
\begin{eqnarray}
  W_1 & = & \frac{1}{\rho} \: \left\{ \: ( C_1 \cos{\varphi} - C_3 \sin{\varphi} )\, 
\u + ( C_2 \cos{\varphi} - C_4 \sin{\varphi} ) \, \v + ( C_5 \sin{\varphi} + C_6
\cos{\varphi} ) \, \w \: \right\} \nonumber \\ 
  W_2 & = & \frac{1}{\rho} \: \left\{ \: ( C_1 \sin{\varphi} + C_3 \cos{\varphi}) \,  \u
+ ( C_2 \sin{\varphi} + C_4 \cos{\varphi} ) \, \v - ( C_5 \cos{\varphi} - C_6
\sin{\varphi} ) \, \w \: \right\} \nonumber \\ 
  W_3 & = & \frac{1}{z} \: \left\{ \: C_7 \, \u - C_8 \, \v + C_9 \, \w \: \right\}  \nonumber \\ 
W_4 & = & \frac{1}{\tau} \: \left\{ \: C_{10} \, \u - C_{11} \, \v + C_{12} \, \w \: \right\}
\nonumber \\
\Phi & = & \frac{v}{\sqrt{2}} \left( \begin{array}{c} - ( H_2+iH_1 )\,e^{-i\varphi} \\ H_0+iH_3 \end{array} \right),
\label{eq:axial} 
\end{eqnarray}
with
\[C_1 = 4\frac{\rho}{X}f_1 \qquad C_2 = 4\frac{\rho z}{X^2}f_2 \qquad C_3 = 4\frac{\rho z}{X^2}f_3 \qquad C_4 = 4\frac{\rho}{X}f_4 \]
\[ C_5 = 4\frac{\rho^2}{X^2}f_5 \qquad C_6 = 4\frac{\rho^2 z}{X^3}f_6 \]
\[ C_7 = 4\frac{\rho z^2}{X^3}f_7 \qquad C_8 = 4\frac{\rho z}{X^2}f_8 \qquad C_9 = 4 \frac{z}{X}f_9 \]
\[C_{10} = 4\frac{\rho \tau}{X^2}f_{10} \qquad C_{11} = 4\frac{\rho  z \tau}{X^3}f_{11} \qquad C_{12} = 4\frac{\tau z}{X^2}f_{12} \]
\beq
H_0 = \frac{z}{X} h_0 \qquad H_1 = 2\frac{\rho z}{X^2}h_1 \qquad H_2 = 2\frac{\rho}{X}h_2 \qquad H_3 = h_3
\label{eq:explicit}
\eeq
and
\[
X^2 \equiv x^2+x_a^2 \equiv \rho^2 + z^2 + \tau^2 +x_a^2 \; ,
\] 
for arbitrary
parameter $x_a$. The axial functions $f_i = f_i(\rho,z,\tau)$ and $h_j =
h_j(\rho,z,\tau)$, with  $i=1,\ldots ,12$ and $j=0,\ldots ,3$, are non-singular
and have reflection symmetry
\begin{eqnarray}
               f_i(\rho,-z,\tau) & = & f_i(\rho,z,\tau) \nonumber \\
               h_j(\rho,-z,\tau) & = & h_j(\rho,z,\tau)\; .
\label{eq:zsym}
\end{eqnarray}
Continuity of the gauge fields at $\rho=0$ demands
\begin{eqnarray}
             f_1(0,z,\tau) & = & f_4(0,z,\tau) \nonumber \\
            -f_2(0,z,\tau) & = & f_3(0,z,\tau)\; .
\end{eqnarray}
All axial functions have furthermore Neumann boundary conditions at $\rho=0$
\begin{eqnarray}
                  \partial_\rho f_i(0,z,\tau) = 0 \nonumber \\
                  \partial_\rho h_j(0,z,\tau) = 0
\end{eqnarray}
and Dirichlet-like boundary conditions at infinity
\beq
            \lim_{|x| \to \infty} \left[ \begin{array}{c} 
               f_1 \\ f_2 \\ f_3 \\ f_4 \\ f_5 \\ f_6 \\ f_7 \\ f_8 \\ f_9 \\
f_{10} \\ f_{11} \\ f_{12} \\ h_0 \\ h_1 \\ h_2 \\ h_3 \end{array} \right] =
\left[ \begin{array}{c} \tau/x \\ -1 \\ (z^2-\rho^2+\tau^2)/x^2 \\ \tau(z^2 -
  \rho^2 + \tau^2)/x^3 \\ 2(\tau^2 + z^2)/x^2 \\ 0 \\ -2\tau/x \\ (\tau^2-
  \rho^2 -
z^2)/x^2 \\ 2 \tau \rho^2 /x^3 \\ (z^2-\rho^2-\tau^2)/x^2 \\ -2\tau/x \\ -2
\rho^2/x^2 \\ 0 \\ -1 \\ -\tau/x \\ (\rho^2 - z^2-\tau^2)/x^2 \end{array} \right].
\label{eq:bcs}
\eeq
The ansatz (\ref{eq:axial}) has, as will be explained in the next
Section, a residual gauge symmetry, which allows for the elimination of
essentially three functions. One possible gauge choice
is the well-known radial gauge $\hat{x}_{\mu} W_{\mu} = 0$ :
\beq
C_1 + C_7 + C_{10} = C_2 - C_8 - C_{11} = C_6 + C_9 + C_{12} = 0\, .
\label{eq:radialgauge}
\eeq
This completes the description of the ansatz\footnote[1]{The $U(1)$ hypercharge
gauge field of
the electroweak standard model can easily be included, just as for the
sphaleron \Sstar ~\cite{K93b}.}. Next, we study its symmetries and explain the
logic behind the construction.    

\section{Symmetries}
The ansatz of the previous Section has, by construction, several symmetries,
which we will now discuss.

\subsection{Axial symmetry}
Generally, scalar and vector fields transform under infinitesimal coordinate
transformations
\beq
      x^\mu \to x^\mu + \delta \alpha \; \xi^\mu(x)
\eeq
as follows:
\alpheqn
\begin{eqnarray}
               \Phi(x) & \to & \Phi(x) + \delta\alpha  \: \L_\xi \Phi(x)
\label{eq:Phitransf} \\
               W_\mu(x) & \to & W_\mu(x) + \delta\alpha \: \L_\xi W_\mu(x)\; ,
\label{eq:Wtransf} 
\end{eqnarray}
\reseteqn
with Lie derivatives
\alpheqn
\begin{eqnarray}
            \L_\xi \Phi & \equiv & \xi^\nu\partial_\nu \Phi 
\label{eq:LiePhi}\\ 
            \L_\xi W_\mu & \equiv & \xi^\nu\partial_\nu W_\mu +
              \left(\partial_\mu \xi^\nu \right) W_\nu\; .
\label{eq:LieW} 
\end{eqnarray}
\reseteqn   
In the present paper we will use a generalized form of invariance of the fields, namely by
allowing for a compensating internal symmetry transformation. 
In short, invariance need only hold up to a gauge transformation (or any other internal
symmetry transformation for that matter). 

The coordinate transformation relevant here is the rotation
\beq
x^\mu \to x^\mu + \delta\varphi\; \xi^\mu_{\rm R} \, ,
\eeq
generated by the vector field
\beq
       \xi_{\rm R}^\mu = (-x_2,x_1,0,0)
\eeq
and infinitesimal parameter $\delta\varphi$. It is then straightforward
to verify the invariance of the ansatz gauge fields (\ref{eq:axial}) under
the transformation (\ref{eq:Wtransf}),
with $\delta\alpha$ replaced by $\delta\varphi$ and $\xi$ by
$\xi_{\rm R}$, combined with the global gauge transformation 
\beq
        W_\mu \to W_\mu - [W_\mu,\tau^3]\; \delta\varphi\; .
\eeq
For the scalar field (\ref{eq:axial}) there is in addition to the
transformation (\ref{eq:Phitransf}),
again with $\delta\alpha$ replaced by $\delta\varphi$ and $\xi$ by
$\xi_{\rm R}$,  a compensating transformation 
\beq
           \Phi \to \Phi - i \, \delta\varphi \left(\frac{\id + \sigma^3}{2}
\right) \Phi\, ,
\eeq
which combines a global $SU(2)$ gauge transformation with a $U(1)$ phase
transformation. The ansatz (\ref{eq:axial}) thus has $U(1)$ invariance
under rotations in the $x_1$,$x_2$ plane.

\subsection{Discrete symmetries}
The ansatz (\ref{eq:axial}--\ref{eq:zsym}) is invariant under the
combined discrete transformation
\beq
       P_{xz} \otimes C \otimes G_{\rm c}\; ,
\label{eq:disc}
\eeq
with $P_{xz}$ the following parity-like transformation of the classical fields
(cartesian coordinates 
$x_1$, $x_2$, $x_3$, $x_4$ being written as $x$, $y$, $z$, $\tau$) : 
\beq
      \left[ \begin{array}{c} W_1 \\ W_2 \\ W_3 \\ W_4 \\ \Phi \end{array}
      \right](x,y,z,\tau) \rightarrow \left[ \begin{array}{r} - W_1 \\ W_2 \\ - W_3 \\
       W_4 \\ \Phi_{\phantom{1}} \end{array} \right] (-x,y,-z,\tau)\; , 
\eeq
$C$ the charge conjugation transformation 
\beq
      \left[ \begin{array}{c} W_\mu   \\ \Phi   \end{array} \right] \rightarrow 
      \left[ \begin{array}{c} W_\mu^* \\ \Phi^* \end{array} \right]   
\eeq
and $G_{\rm c}$ the global $SU(2)$ gauge transformation with gauge parameter function 
$\Gamma = -\id_{2}$ in the center of the group
\beq
      \left[ \begin{array}{c} W_\mu \\ \Phi \end{array} \right] \rightarrow 
      \left[ \begin{array}{c} W_\mu \\ - \Phi \end{array} \right]  \; . 
\eeq

The second discrete symmetry of the ansatz acts only on the fields in the $z =
0$ plane and corresponds to the usual parity transformation $P$
\beq
\left[ \begin{array}{c} W_1 \\ W_2 \\ W_4 \\ \Phi \end{array}
\right](x,y,0,\tau) \rightarrow \left[ \begin{array}{c} -W_1 \\ -W_2 \\ -W_4 \\ \Phi
\end{array} \right](-x,-y,0,-\tau)\; ,
\eeq
provided the ansatz functions (\ref{eq:explicit}) obey the
following conditions :
\begin{eqnarray}
-f_1(\rho,0,-\tau) &=& f_1(\rho,0,\tau) \nonumber \\
-f_4(\rho,0,-\tau) &=& f_4(\rho,0,\tau) \nonumber \\
f_5(\rho,0,-\tau) &=& f_5(\rho,0,\tau) \nonumber \\
f_{10}(\rho,0,-\tau) &=& f_{10}(\rho,0,\tau) \nonumber \\
-h_2(\rho,0,-\tau) &=& h_2(\rho,0,\tau) \nonumber \\
h_3(\rho,0,-\tau) &=& h_3(\rho,0,\tau)\; .
\label{eq:parity}
\end{eqnarray}
Note that the same parity reflection symmetry holds for the sphaleron
\Sstar ~\cite{K93b}, which corresponds in fact to the $z=0$ slice of \Istar.

Both discrete symmetries also distinguish the configuration at the ``top''
of the NCL constructed previously \cite{K93a}. The crucial idea
behind the ansatz of Section 3 was to generalize that particular NCL configuration,
while respecting the axial and discrete symmetries present. 
Further details can be found in Appendix~A. 

\subsection{Residual gauge symmetry}
The axisymmetric ansatz (\ref{eq:axial}) is form invariant under gauge
transformations
\begin{eqnarray}
       W_\mu & \to & \Gamma \left(W_\mu + \partial_\mu \right) \Gamma^{-1} \nonumber \\
       \Phi & \to & \Gamma \, \Phi \, ,
\label{eq:resgauge}
\end{eqnarray}
with gauge parameter function
\beq
         \Gamma = \exp\left( \omega_1 \, \u + \omega_2 \, \v + \omega_3 \, \w
\right),
\label{eq:gaugemap}
\eeq
where $\omega_a = \omega_a(\rho,z,\tau)$.  In order to maintain the discrete
symmetries of the ansatz, there are the following conditions :
\begin{eqnarray}
\omega_1(\rho,-z,\tau) &=& \omega_1(\rho,z,\tau) \nonumber \\
-\omega_2(\rho,-z,\tau) &=& \omega_2(\rho,z,\tau) \nonumber \\
-\omega_3(\rho,-z,\tau) &=& \omega_3(\rho,z,\tau) \nonumber \\
-\omega_1(\rho,0,-\tau) &=& \omega_1(\rho,0,\tau) \, .
\label{eq:resgaugeomegasymm}
\end{eqnarray}
The explicit transformations of the
coefficient functions $C_i$ and $H_j$ are somewhat involved and are given in
Appendix B.
As mentioned above, these residual gauge transformations eliminate
essentially 3 functions from the ansatz.

\section{Action}
The ansatz of section 3 gives for the Yang-Mills-Higgs action 
\beq
      A_{\mbox {\rm \tiny YMH}} = 4\pi \int\limits^{\infty}_{-\infty}\!d\tau
\int\limits_0^\infty \!dz  \int\limits_0^\infty \!d \rho  \, \rho \, a_{\mbox {\rm \tiny YMH}}\, ,
\label{eq:AYMHansatz}
\eeq
with actiondensity
\beq
       a_{\mbox {\rm \tiny YMH}} = a_{ \mbox {\rm \tiny WKIN}} + a_{ \mbox {\rm
\tiny HKIN}} + a_{\mbox {\rm \tiny HPOT}} \, ,
\label{eq:adensity}
\eeq
where
\begin{eqnarray}
& & a_{\mbox{\tiny WKIN}} =  \nonumber \\
& & \nonumber \\
& \displaystyle \frac{1}{2 g^2} \Bigg\{ & \left(\frac{\D_z C_1}{\rho}-\frac{\D_\rho
C_7}{z}-\frac{C_2C_9}{\rho z}-\frac{C_6C_8}{\rho z}\right)^2+\left(\frac{\D_\tau C_1}{\rho}-\frac{\D_\rho
C_{10}}{\tau}-\frac{C_2C_{12}}{\rho\tau}-\frac{C_6C_{11}}{\rho
\tau}\right)^2 \nonumber \\
&+& \left(\frac{\D_z C_2}{\rho}+\frac{\D_\rho C_8}{z}+\frac{C_1C_9}{\rho
z}-\frac{C_6C_7}{\rho z}\right)^2+\left(\frac{\D_\tau C_2}{\rho}+\frac{\D_\rho
C_{11}}{\tau}+\frac{C_1C_{12}}{\rho \tau}-\frac{C_6C_{10}}{\rho
\tau}\right)^2 \nonumber \\
&+& \left(\frac{\D_zC_6}{\rho}-\frac{\D_\rho C_9}{z}+\frac{C_1C_8}{\rho z}+\frac{C_2C_7}{\rho
z}\right)^2 +
\left(\frac{\D_\tau
C_6}{\rho}-\frac{\D_\rho
C_{12}}{\tau}+\frac{C_1C_{11}}{\rho\tau}+\frac{C_2C_{10}}{\rho\tau}\right)^2
\nonumber \\
&+&\left(\frac{\D_\tau C_7}{z}-\frac{\D_z C_{10}}{\tau}+\frac{C_8C_{12}}{z
  \tau}-\frac{C_9C_{11}}{z\tau}\right)^2 + \left(\frac{\D_\tau
C_8}{z}-\frac{\D_z C_{11}}{\tau}-\frac{C_7 C_{12}}{z
\tau}+\frac{C_9C_{10}}{z \tau}\right)^2 \nonumber \\
&+& \left(\frac{\D_\tau
C_9}{z}-\frac{\D_z
C_{12}}{\tau}+\frac{C_7 C_{11}}{z\tau}-\frac{C_8C_{10}}{z \tau}\right)^2 
+ \left(\frac{\D_\rho
  C_3}{\rho}-\frac{C_4C_6}{\rho^2}-\frac{C_2}{\rho^2}\left(C_5-1\right)\right)^2 \nonumber \\
&+&\left(\frac{\D_z
C_3}{\rho}-\frac{C_4 C_9}{\rho
z}+\frac{C_8}{\rho z}\left(C_5-1\right) \right)^2 + \left(\frac{\D_\tau
C_3}{\rho}-\frac{C_4C_{12}}{\rho \tau}+\frac{C_{11}}{\rho
\tau}\left(C_5-1\right)\right)^2 \nonumber \\
&+& \left(\frac{\D_\rho
C_4}{\rho}+\frac{C_3C_6}{\rho^2}+\frac{C_1}{\rho^2}\left(C_5-1\right)\right)^2
 + \left(\frac{\D_z C_4}{\rho}+\frac{C_3C_9}{\rho z}+\frac{C_7}{\rho
   z}\left(C_5-1\right)\right)^2 \nonumber \\
&+& \left(\frac{\D_\tau
C_4}{\rho}+\frac{C_3C_{12}}{\rho\tau}+\frac{C_{10}}{\rho
\tau}\left(C_5-1\right)\right)^2 + \left(\frac{\D_\rho
C_5}{\rho}+\frac{C_2C_3}{\rho^2}-\frac{C_1C_4}{\rho^2}\right)^2 \nonumber \\
&+& \left.\left(\frac{\D_zC_5}{\rho}-\frac{C_3C_8}{\rho
z}-\frac{C_4C_7}{\rho z}\right)^2 +\left(\frac{\D_\tau
C_5}{\rho}-\frac{C_3C_{11}}{\rho\tau}-\frac{C_4C_{10}}{\rho\tau}\right)^2 \right\} 
\label{eq:awkin}
\end{eqnarray}

\begin{eqnarray}
& & a_{\mbox{\tiny HKIN}} = \nonumber \\
& & \nonumber \\
& \displaystyle \frac{v^2}{2}\Bigg\{ & \left(\D_\rho
H_0-\frac{H_1C_1}{2\rho}-\frac{H_2C_2}{2\rho}-\frac{H_3C_6}{2\rho} \right)^2+ \left(\D_z H_0-\frac{H_1C_7}{2z}+\frac{H_2C_8}{2z}-\frac{H_3C_9}{2z}
\right)^2 \nonumber \\
&+& \left(\D_\tau H_0
-\frac{H_1C_{10}}{2\tau}+\frac{H_2C_{11}}{2\tau}-\frac{H_3C_{12}}{2\tau} \right)^2
+\left( \D_\rho H_1
+\frac{H_0C_1}{2\rho}-\frac{H_2C_6}{2\rho}+\frac{H_3C_2}{2\rho}\right)^2
\nonumber \\
&+& \left(\D_z H_1 + \frac{H_0C_7}{2z}-\frac{H_2C_9}{2z}-\frac{H_3C_8}{2z}
\right)^2+\left(\D_\tau H_1
+\frac{H_0C_{10}}{2\tau}-\frac{H_2C_{12}}{2\tau}-\frac{H_3C_{11}}{2\tau}
\right)^2 \nonumber \\
&+& \left(\D_\rho H_2 +
\frac{H_0C_2}{2\rho}+\frac{H_1C_6}{2\rho}-\frac{H_3C_1}{2\rho} \right)^2+ \left( \D_z H_2 -\frac{H_0C_8}{2z}+\frac{H_1C_9}{2z}-\frac{H_3C_7}{2z}
\right)^2 \nonumber \\
&+& \left(\D_\tau H_2
-\frac{H_0C_{11}}{2\tau}+\frac{H_1C_{12}}{2\tau}-\frac{H_3C_{10}}{2\tau}
\right)^2 +\left(\D_\rho H_3
  +\frac{H_0C_6}{2\rho}-\frac{H_1C_2}{2\rho}+\frac{H_2C_1}{2\rho} \right)^2
   \nonumber \\
& + & \left(\D_z H_3+\frac{H_0C_9}{2z}+\frac{H_1C_8}{2z}+\frac{H_2C_7}{2z}
\right)^2 + \left( \D_\tau H_3
  +\frac{H_0C_{12}}{2\tau}+\frac{H_1C_{11}}{2\tau}+\frac{H_2C_{10}}{2\tau}   
  \right)^2  \nonumber \\
&+& \left(
\frac{H_0C_4}{2\rho}+\frac{H_1}{\rho}-\frac{H_1C_5}{2\rho}-\frac{H_3C_3}{2\rho}
\right)^2+
\left(\frac{H_0C_3}{2\rho}-\frac{H_2}{\rho}+\frac{H_2C
    _5}{2\rho}+\frac{H_3C_4}{2\rho}\right)^2 
  \nonumber \\
&+& \left. \left(\frac{H_0C_5}{2\rho}+\frac{H_1C_4}{2\rho}-\frac{H_2C_3}{2\rho}
\right)^2+
\left(\frac{H_1C_3}{2\rho}+\frac{H_2C_4}{2\rho}-\frac{H_3C_5}{2\rho}\right)^2\right\}
\label{eq:ahkin} \\
a_{\mbox{\tiny HPOT}} & = & \lambda\, \frac{v^4}{4}
\left(\, H_0^2+H_1^2+H_2^2+H_3^2 - 1 \, \right)^2.
\label{eq:ahpot}
\end{eqnarray}
For brevity, we have not made the functions $C_i$ and $H_j$ explicit by inserting
(\ref{eq:explicit}), 
but if one does one readily verifies that the action
density is finite everywhere and vanishes at infinity. In addition, we can fix
the gauge with conditions (\ref{eq:radialgauge}).

The constraint operator (\ref{eq:ac}, \ref{eq:o8}) gives for the ansatz fields (\ref{eq:axial}) 
a term in the total action of the form
\beq
     A_{\mbox {\rm \tiny C}} = \frac{\kappa}{g^2} \left( \frac{4\pi}{M_W^4}
\int\limits^{\infty}_{-\infty}\!d\tau
\int\limits_0^\infty \!dz  \int\limits_0^\infty \!d \rho\,\rho\:q_{\rm P}^2
- \frac{8\pi^2c}{(M_{\rm W}\varrho)^4} \right)\, ,
\label{eq:ACansatz}
\eeq
with Pontryagin density $q_{\rm P}$ given by

\newcommand{\DLR}{\stackrel{\leftrightarrow}{D}}
\newcommand{\DR} {\stackrel{\rightarrow}{\partial}}
\begin{eqnarray}
\rho^2 z\tau\; q_{\rm P} &=&
    C_1\DLR_{\tau z} C_3  + C_2\DLR_{\tau z} C_4   + C_5\DLR_{\tau z} C_6
  + C_7\DLR_{\rho \tau} C_3 \nonumber \\
&+& C_4\DLR_{\rho \tau} C_8+ C_5  \DLR_{\rho \tau} C_9
  + C_3\DLR_{\rho z} C_{10}+ C_{11} \DLR_{\rho z} C_4+ C_{12}\DLR_{\rho z}C_5
  \nonumber \\
&+& \rho\DR_\rho\big[\left(C_5-1\right)\left(C_7C_{11}-C_8C_{10}\right)+C_9\left(C_3C_{11}+C_4C_{10}\right)
  - C_{12}\left(C_3C_8+C_4C_7\right)\big] \nonumber \\
&+& z\DR_z\big[\left(1-C_5\right)\left(C_2C_{10}+C_1C_{11}\right) -
    C_6\left(C_3C_{11} + C_4C_{10} \right)
  + C_{12}\left(C_1C_4 - C_2C_3\right)\big] \nonumber\\
&+& \tau\DR_\tau\big[\left(C_5 - 1\right)\left(C_1C_8+C_2C_7\right)+C_6\left(C_3C_8+C_4C_7\right)
  - C_9\left(C_1C_4-C_2C_3\right)\big]\; , \nonumber \\
\label{eq:qp}
\end{eqnarray}
in terms of
\begin{eqnarray*}
\DLR_{\tau z} &\equiv &
\stackrel{\leftarrow}{\D}_{\tau} \tau  z   \stackrel{\rightarrow}{\D}_{z}  -
\stackrel{\leftarrow}{\D}_{z}    z  \tau   \stackrel{\rightarrow}{\D}_{\tau}  \\
\end{eqnarray*}
and similarly for  $\DLR_{\rho \tau}$ and $ \DLR_{\rho z}$, 
where the partial derivatives $\stackrel{\leftarrow}{\D}$ and 
$\stackrel{\rightarrow}{\D}$ operate
to the left and to the right, respectively.
Again, we have to insert (\ref{eq:explicit})
and implement the gauge fixing conditions (\ref{eq:radialgauge}).

It is possible to get more compact expressions for the Pontryagin density
and actiondensities, see Appendix C.
With (\ref{eq:AYMHansatz}, \ref{eq:ACansatz}) one then obtains the final expression for
the total action (\ref{eq:A}--\ref{eq:c}) evaluated with the ansatz
fields. Furthermore, the Pontryaginindex (\ref{eq:Qp}) can be shown to
vanish for the fields of the ansatz.
This can be established most easily by use of the Chern-Simons current
\beq
       j_\mu \equiv -\epsilon_{\mu\nu\alpha\beta} \, {\rm Tr}
\,\left[W_\nu\left(\D_\alpha W_\beta +
                  \frac{2}{3} \, W_\alpha W_\beta \right)\right],
\eeq
whose divergence gives the Pontryagin density $\D_\mu j_\mu = q_{\rm P}$,
so that
\[
  Q_{\rm P} = \frac{1}{8\pi^2} \oint\limits_{S^3_{\infty}}\! d\Sigma_\mu \, j_\mu\; .
\]
For the ansatz fields at infinity (\ref{eq:axial}, \ref{eq:explicit},
\ref{eq:bcs}) the Chern-Simons current vanishes identically and
\beq
         Q_{\rm P}(W_{{\rm I}^{\star}}) = 0\; .
\eeq
This result for the topological charge $Q_{\rm P}$
also follows from the simple observation that the
solution \Istar ~lies  on a {\em continuous} path of configurations (NCL)
connected to the vacuum configuration $W_{\rm vac} = 0$,
which is topologically trivial $Q_{\rm P}(0) = 0$.

\section{Solution}
The ansatz (\ref{eq:axial}--\ref{eq:zsym}) 
is self-consistent, which means that the field equations
(from $\delta A/\delta W = \delta A / \delta \Phi = 0$) reduce to 16 equations,
which are precisely equal to those obtained from variations $\delta C_i$ and $\delta
H_j$ of the ansatz action (\ref{eq:AYMHansatz}, \ref{eq:ACansatz}). This agrees
with the so-called principle of symmetric criticality \cite{P79}, which states
that in the quest of stationary points it suffices,
under certain conditions, to consider variations that respect
the symmetries of the ansatz (rotation and reflection symmetries in our case).

An analytic solution of the resulting non-linear partial differential equations for
the functions $f_i(\rho,z,\tau)$ and $h_j(\rho,z,\tau)$ seems to be impossible
and even an accurate numerical evaluation is a major enterprise, which we have
to postpone for the moment. Still, it is possible to argue that there does exist a
non-trivial solution, i.~e. a solution different from the vacuum ($W=0$, $|\Phi|^2
= \frac{1}{2} v^2$) in whatever complicated gauge. The crucial observation is that
the boundary conditions and symmetries of the ansatz imply the existence of at
least one point where the total Higgs field vanishes $\Phi = 0$,
which is a gauge invariant statement.

  \begin{figure}[p]
  \centerline{\psfig{figure=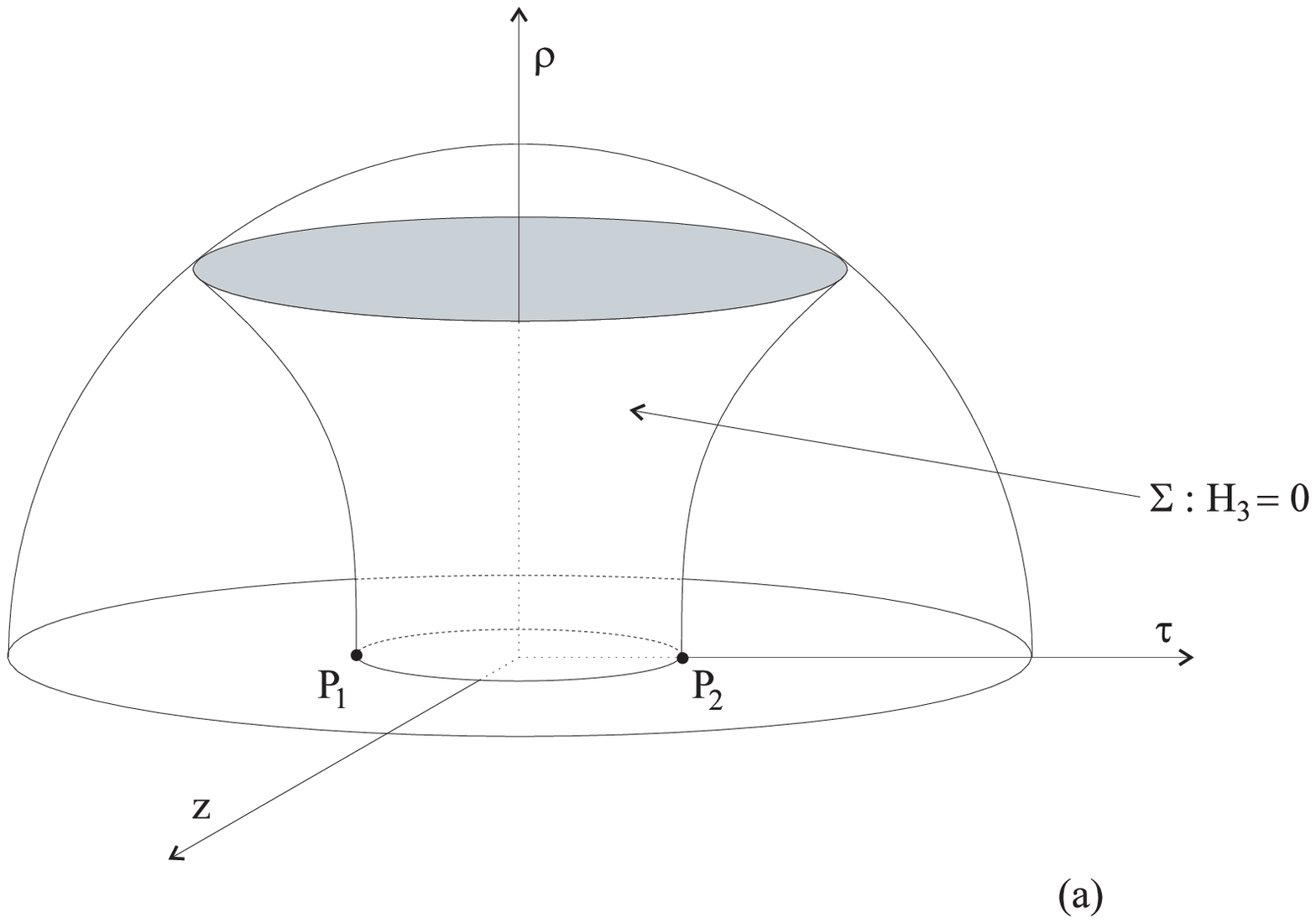,width=10cm,bbllx=2154pt,bblly=2281pt,bburx=2605pt,bbury=2599pt}}
  \vspace{1.5cm}
  \centerline{\psfig{figure=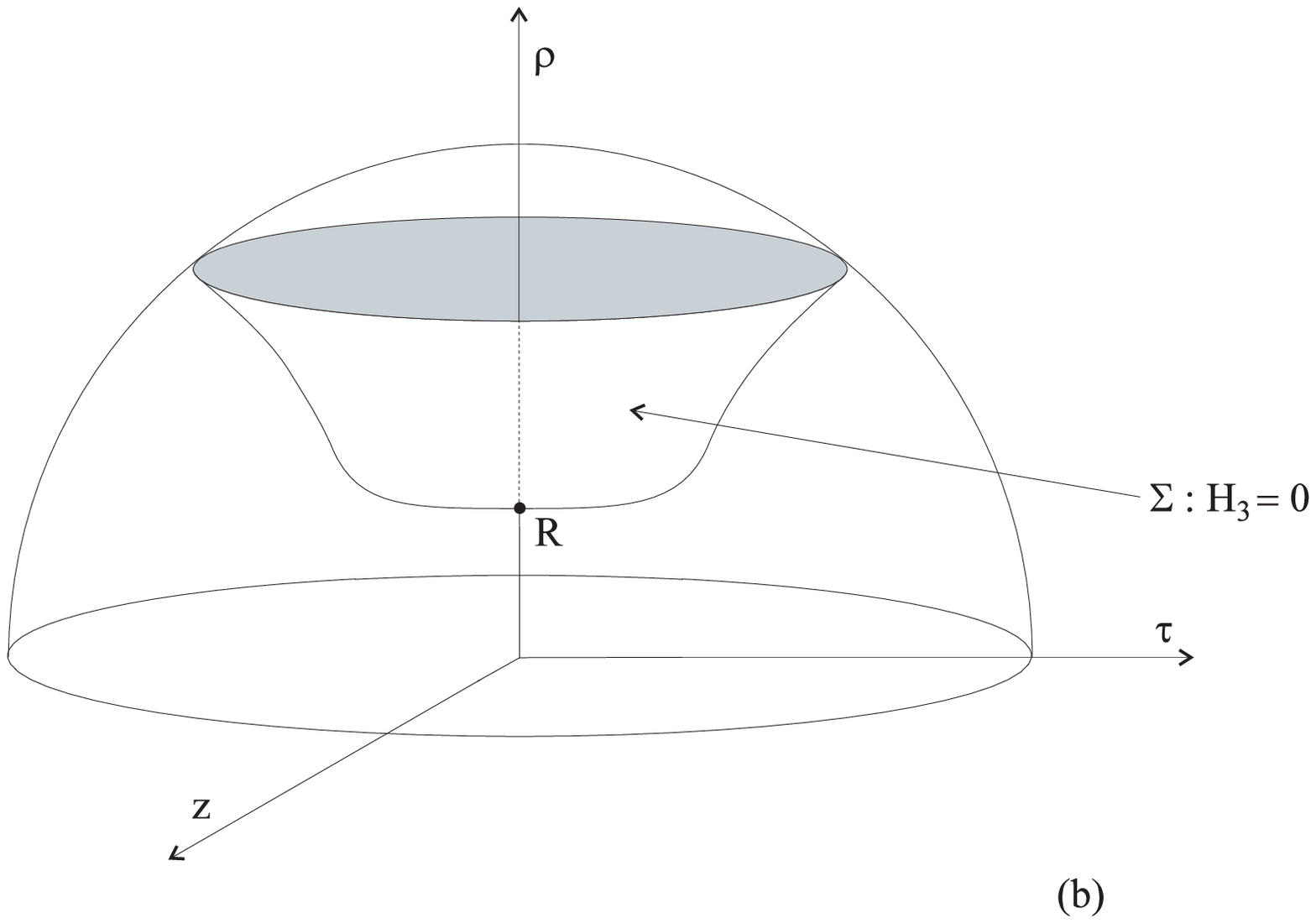,width=10cm,bbllx=2154pt,bblly=2279pt,bburx=2605pt,bbury=2599pt}}
  \caption{Sketch of the $H_3=0$ surface $\Sigma$ as it propagates in from
   the sphere at infinity.  {\em A priori\/} there are  two possibilities :
   (a) a single connected surface embedded in $\R^4$, with Higgs field vanishing at, generically,
   two points $P_1$ and $P_2\,$; (b) a ``disconnected'' surface in $\R^4$, with Higgs field vanishing at
   $R$, which corresponds to a ring of zeros in $\R^4$.}
  \end{figure}

Consider  then the $H_3 = 0$ surface $\Sigma$ as it comes in from infinity
(far out $\Sigma$ is, according to (\ref{eq:bcs}), given by the equation $z^2+\tau^2 = \rho^2\,$).
The simplest possibility is that of a single, connected surface as shown in
Fig. 1a.  The ansatz (\ref{eq:explicit}) gives, in that case, at least one point
on the $\tau$-axis with {\em all} coefficient functions $H_j$ vanishing, i.~e. $\Phi = 0$
there. Alternatively, the $H_3 = 0$ surface could ``pinch off'', as shown in
Fig. 1b, in which case the ansatz (\ref{eq:explicit}, \ref{eq:parity})
gives a ring of zeros for the Higgs field. Whichever possibility is realized is for
the field equations to decide. The same two basic possibilities occur for the
static sphaleron \Sstar, where the numerical solution of the field
equations appears to indicate the first alternative \cite{K93b}. Hence,
one expects also \Istar ~to resemble a di-atomic molecule, albeit on a
scale of $M_{\rm W}^{-1} \sim 10^{-8} {\rm {\AA}}$.

\section{Discussion}
In this paper we have presented a selfconsistent ansatz for a new constrained
instanton \Istar ~in $SU(2)$ \YMHth. Further work on the numerical solution of the
reduced field equations remains to be done, but we can already make some
general remarks, see also \cite{K93a,K93b}.

The constrained instanton \Istar ~may be thought of as locating the top of the
action barrier for the global $SU(2)$ anomaly \cite{W82}.
The electroweak standard model has, of course,
no global $SU(2)$ anomaly, the total number of left-handed
fermion doublets being even.
Still, the barrier structure in configuration space and the corresponding
constrained instanton solution remain.

A more direct application of \Istar ~to the electroweak interactions may 
be the asymptotics of Feynman perturbation theory.
From the outset it is important to realize \cite{V94} that the
\ew ~\sm ~has only been established in (low order) perturbation theory.
High-order contributions $c_n\, g^{2n}$ to an arbitrary physical observable
may be estimated \cite{L77} by saddle-point approximation of
the euclidean path integral. It appears that \Istar
~provides the relevant saddle-point. In particular, the \Istar ~negative mode
(which projects onto the \ncl) plays a important role. Assuming there to be a
single negative mode and assuming the integral over the collective
coordinate $\varrho$ to be dominated by the value $\varrho = 0$, the
expected asymptotic behavior is
\beq
 c_n \,g^{2n} \sim \frac{n!}{ \left( A_{\mbox {\tiny YMH}} ^\star (0)\right)^n}
 \sim \frac{n!}{ \left( 16 \, \pi^2 \right)^n}\; g^{2n}\; ,
\eeq
where for the instanton action $A_{\mbox {\tiny YMH}}^\star (\varrho) = 16 \pi^2/g^2
+ {\rm O}(\varrho^2 v^2)$ is used \cite{K93a}. If this is indeed the asymptotic behaviour
of standard \ew ~perturbation theory, then the series is not even Borel
summable. Physics considerations (such as causality and unitarity) 
should tell us how to make sense of perturbation theory or, more generally, how to define
electroweak field theory non-perturbatively.

\begin{appendix}
\renewcommand{\theequation}{A.\arabic{equation}}
\setcounter{equation}{0}

\section*{Appendix A\label{sec:ncl}}
Here, we review the \ncl ~(NCL) given in \cite{K93a} and put it in the
appropriate form for the construction of the ansatz of Section 3.

The basic structure of this loop of 4-dimensional configurations is to first
create and separate (to a distance $d\,$) an BPST-like instanton--anti-instanton
pair, then make a complete relative isospin rotation, and finally collapse and
annihilate the pair.
For the present purpose we only need the isospin rotation part of
the NCL with loop-parameter $\omega \in [-\pi/2,+\pi/2]$, whereas the whole NCL
has $\omega$ running from $-3\pi/2$ to $+3\pi/2$. 

With the same notation as in
Section 3 and defining the radial coordinate $r^2 \equiv \rho^2 + z^2$, the
configurations of the NCL are for $\omega \in [-\pi/2, +\pi/2]$
\begin{eqnarray}
               W_\mu & = & -f\; \D_\mu U \,\; U^{-1} \nonumber \\
               \Phi & = & \frac{v}{\sqrt{2}}\; h \; U \, \vectwo{0}{1},
\label{eq:ncl}
\end{eqnarray}
with the $SU(2)$ matrix\footnote[1]{the functions $g_\pm$ of \cite{K93a}
are set to unity and a global gauge transformation  (\ref{eq:resgauge})
with $\Gamma=i\sigma_3$ has been performed.}
\beq
           U = i\sigma_3\;  e^{(\omega + \pi/2)i\sigma_3}\left(\hat{x}_- \cdot \sigma
          \right) e^{-(\omega+\pi/2)i\sigma_3}\left(\hat{x}_+ \cdot \sigma \right)^\dagger
\eeq
and definitions
\begin{eqnarray*}
                   \hat{x}^\mu_\pm & \equiv & x_\pm^\mu / | x_\pm | \\ 
                    x^\mu_\pm & \equiv & \left( x^1,x^2,x^3,x^4 \pm d/2
                                                          \right) \\ 
                    \sigma^\mu & \equiv & \left( i\sigma^1,i\sigma^2,i\sigma^3,
                                                                \id_2 \right).
\end{eqnarray*} 
The axial functions $f = f(r,\tau)$ and $h = h(r,\tau)$
have boundary conditions\footnote[7]{ the additional conditions
$f(r,-\tau)  =  f(r,\tau)$ and  $h(r,-\tau)  =  h(r,\tau)$
of \cite{K93a} are, strictly speaking, not essential.}
\[ 
       f(0,\pm d/2) = h(0,\pm d/2) = 0
\]
\beq
        \lim\limits_{|x| \to \infty} f =  \lim\limits_{|x| \to \infty} h = 1 \, .
\eeq
This completes our review of the crucial part of the NCL, which is based on the
topologically non-trivial loop of mappings $U(\omega)$ of the sphere at infinity
$S^3_{\infty}$ into the group manifold $SU(2) = S^3$.

The NCL configurations (\ref{eq:ncl}) are axially symmetric and in the notation of
(\ref{eq:axial}) the corresponding coefficient functions $C_i$ and $H_j$ are given by
\begin{eqnarray*}
C_1 &=& \frac{\rho f}{x_+^2 x_-^2} \left\{\left( 
\left(2\tau-d\right)\left(\tau^2-\frac{d^2}{4}+z^2\right)-2 z^2d\right)
-2zx_+^2\sin{2\omega} \right. \\ \nonumber
& & +\left. \left(2\tau-d\right)x_+^2 \cos{2\omega}
-2 z \rho^2\sin{4\omega}
+\rho^2\left(2\tau+d\right) \cos{4\omega}
\phantom{\frac{1}{1}} \!\!\right\}  \\& & \\
C_2 &=& -\frac{\rho f}{x_+^2 x_-^2} \left\{ 2z\left(\tau d+\tau^2-\frac{3}{4}\, d^2 +z^2\right) +\left(2\tau-d\right)x_+^2 \sin{2\omega} \right. \\ \nonumber
& & \left. +\, 2 z x_+^2 \cos{2\omega} + \rho^2\left(d+2\tau\right)\sin{4\omega}+2z\rho^2\cos{4\omega}\phantom{\frac{1}{1}}\!\!\right\} \\ & & \\
C_3 &=& \frac{\rho f}{x_+^2 x_-^2} \left\{ 2z\left(\tau
d+\tau^2-\frac{3}{4}\, d^2+z^2 \right) -\left(2\tau-d\right)\left(2\rho^2-x_+^2\right)\sin{2\omega} \right. \\
\nonumber 
& & -\left. 2z\left(2\rho^2-x_+^2\right)\cos{2\omega} -\rho^2\left(2\tau+d\right)\sin{4\omega}-2z\rho^2\cos{4\omega}\phantom{\frac{1}{1}}
\!\!\right\} \\& & \\
C_4 &=& \frac{\rho f}{x_+^2 x_-^2} \left\{\left(
\left(2\tau-d\right)\left(\tau^2-\frac{d^2}{4}+z^2\right)-2 z^2d\right)
+2z\left(2\rho^2-x_+^2\right)\sin{2\omega} \right. \\ \nonumber
& & \left. -\left(2\tau-d\right)\left(2\rho^2-x_+^2\right)\cos{2\omega} + 2z\rho^2\sin{4\omega}-\rho^2\left(2\tau+d\right)\cos{4\omega}\phantom{\frac{1}{1}}
\!\!\right\}  \\& & \\
C_5 &=& \frac{4\rho^2 f}{x_+^2 x_-^2} \left\{\left( \tau^2 + z^2 +\frac{d^2}{4}\right) +
dz\sin{2\omega} +\left(\tau^2 + z^2 -\frac{d^2}{4}\right)\cos{2\omega}\phantom{\frac{1}{1}}
\!\!\right\}  \\& & \\
C_6 &=& \frac{4\rho^2 f}{x_+^2 x_-^2}
\left\{\left(\frac{d^2}{4}-z^2 -\tau^2\right) \sin{2\omega} +dz\cos{2\omega} \phantom{\frac{1}{1}}
\!\!\right\}\\ &&\\
C_7 &=& -2\frac{\rho z f}{x_+^2 x_-^2}\left\{
z\left(2 \tau -d \right) -\left( \tau d + z^2 +
\rho^2 - \tau^2 + \frac{3}{4}\, d^2 \right) \sin{2\omega}
\right. \nonumber \\
& &+ \left. z\left(d + 2 \tau  \right) \cos{2\omega} - \rho^2
\sin{4\omega}  \phantom{\frac{1}{1}}
\!\!\right\} \\& & \\
C_8 &=& 2\frac{\rho z f}{x_+^2 x_-^2}\left\{
\left(\tau^2 -z^2 -\tau d + \frac{d^2}{4} \right) -\left(\tau d + z^2 + \rho^2 - \tau^2 + \frac{3}{4} \, d^2
\right) \cos{2\omega} \right. \nonumber \\ 
& & - \left. z\left(2\tau  + d \right) \sin{2\omega} - \rho^2
\cos{4\omega}  \phantom{\frac{1}{1}}
\!\!\right\} \\& & \\
C_9 &=& 2\frac{z f}{x_+^2 x_-^2}\left\{
\left(2 \rho^2 \tau + \tau^2 d -z^2 d -
\frac{d^3}{4}\right) +2z\rho^2 \sin{2\omega}
+\rho^2\left( 2\tau -d \right) \cos{2\omega}
 \phantom{\frac{1}{1}}
\!\!\right\} \\& & \\
C_{10} &=& 2 \frac{\rho \tau f}{x_+^2 x_-^2}\left\{
\left( z^2 -\tau^2 + \tau d - \frac{d^2}{4}  \right) -\left(\rho^2 + \tau d - z^2 +
\frac{d^2}{4} + \tau^2 \right) \cos{2\omega}
\right. \nonumber \\ 
& & + \left. z\left(2 \tau  - d \right) \sin{2\omega} - \rho^2 \cos{4\omega}
\phantom{\frac{1}{1}}  \right\} \\& & \\
C_{11} &=& -2\frac{\rho \tau f}{x_+^2x_-^2} \left\{
z\left(2 \tau -d\right) +\left( \rho^2 + \tau d -z^2 +\frac{d^2}{4} +
\tau^2 \right) \sin{2 \omega} \right. \\
& & \left.  + \, z\left(  2 \tau - d \right) \cos{2\omega} + \rho^2 \sin{4\omega}
\phantom{\frac{1}{1}} \right\}
\\& & \\
C_{12} &=& -4 \frac{f \tau}{x_+^2x_-^2} \left\{
z\left(\tau d  + \rho^2
\right) +\rho^2\left( \frac{d}{2}-\tau 
\right) \sin{2 \omega} + \rho^2 z \cos{2 \omega}   \right\} \\ &&\\
H_0 &=& \frac{h}{x_+x_-} \left\{-zd+\rho^2 \sin{2\omega} \right\}  \\& & \\
H_1 &=&-\frac{h}{x_+x_-} \rho \left\{
z +\left( \tau + \frac{d}{2} \right)
\sin{2\omega} + z \cos{2\omega}  \right\}  \\& & \\
H_2 &=&-\frac{h}{x_+x_-}\rho \left\{
\left(\tau - \frac{d}{2}\right) - z \sin{2\omega} + \left(\tau + \frac{d}{2}
\right) \cos{2\omega}   \right\}  \\& & \\
H_3 &=& \frac{h}{x_+
x_-}\left\{\left(\frac{d^2}{4}-z^2-\tau^2\right) +\rho^2\cos{2\omega}\right\}.
\end{eqnarray*}
\beq \label{eq:axncl} \eeq

\noindent Only for $\omega = 0$ (and $\omega = \pm \pi/2$) do we have the discrete
symmetries described in Section 4.2. In $H_0$, for example, there is a term
proportional to $\rho^2 \sin{2\omega}$ that breaks the discrete symmetry
(\ref{eq:disc}), whereas the other term proportional to $zd$ respects it.

The ansatz of Section 3 then is a generalization of the configuration (\ref{eq:axncl})
at the top ($\omega =0$) of this particular
NCL, keeping the axial and discrete symmetries present and taking over the
boundary conditions at infinity.
In fact, the ansatz
(\ref{eq:axial}, \ref{eq:explicit}) has {\em a priori\/} 16 independent functions
$f_i(\rho,z,\tau)$ and $h_j(\rho,z,\tau)$, whereas the maximal configuration of
the NCL (\ref{eq:ncl}) has only two, namely $f(r,\tau)$ and $h(r,\tau)$. This
illustrates the degree of generalization required to obtain an exact
solution of the field equations.  

\renewcommand{\theequation}{B.\arabic{equation}}
\setcounter{equation}{0}
\section*{Appendix B}

Here, we list the transformation properties of the coefficient functions
$C_i$, $H_j$ of the ansatz (\ref{eq:axial}) under the residual gauge symmetries
(\ref{eq:resgauge}, \ref{eq:gaugemap}).

The transformations of the gauge field
coefficients functions $C_i$ ($i=1,\ldots,12$) are the following:
\beq
\begin{array}{cccl}
\left( \begin{array}{c} C_1 \\ C_2 \\ C_6  \end{array} \right)
& \to &
\left(\begin{array}{c} T_{\rho1}\\T_{\rho2}\\T_{\rho3}\end{array}\right)
& +\; R
\left( \begin{array}{c} C_1 \\ C_2 \\ C_6 \end{array} \right) \\ & & & \\ 
\left( \begin{array}{c} C_3 \\ C_4 \\ 1-C_5 \end{array} \right) & \to & &\; R
\left( \begin{array}{c} C_3 \\ C_4 \\ 1-C_5 \end{array} \right) \\ & & & \\
\left( \begin{array}{c} C_7 \\ - C_8 \\ C_9 \end{array} \right)
&\to &
\left(\begin{array}{c} T_{z 1}\\T_{z 2}\\T_{z 3}\end{array}\right)
& +\; R\left(
\begin{array}{c} C_7 \\ -C_8 \\ C_9 \end{array} \right) \\ & & & \\
\left( \begin{array}{c} C_{10} \\ -C_{11} \\ C_{12} \end{array} \right)
& \to &
\left(\begin{array}{c} T_{\tau1}\\T_{\tau2}\\T_{\tau3}\end{array}\right)
& +\; R \left( \begin{array}{c} C_{10} \\ -C_{11} \\ C_{12} \end{array} \right),   
\end{array}
\label{ctransf}
\eeq
with the $SO(3)$ matrix
\[
    R \equiv \left( \begin{array}{ccc} R_{11} & R_{12} & R_{13} \\
                                       R_{21} & R_{22} & R_{23} \\
                                       R_{31} & R_{32} & R_{33}
                    \end{array} \right)\, ,
\]
defined in terms of
\begin{eqnarray*}
 R_{ca} & \equiv & \frac{1}{\Omega^2} \left[\,\omega_c \omega_a \left(1 -
\cos{\Omega} \right) - \epsilon_{cab} \omega_b \Omega \sin{\Omega} +
\delta_{ca} \Omega^2 \cos{\Omega} \,\right] \\
 T_{\bar{\alpha} c} &\equiv& 
\frac{1}{\Omega^2} \left(\bar{\alpha}\, \D_{\bar{\alpha}}\, \omega_a
\right) \left[\, \omega_c \omega_a \left( \frac{\sin{\Omega}}{\Omega}- 1 \right)
+ \epsilon_{cab} \omega_b \left(1 - \cos{\Omega} \right)-\delta_{ca} \Omega
\sin{\Omega} \,\right] \\
\Omega^2 & \equiv& \omega_1^2 + \omega_2^2 + \omega_3^2\; ,
\end{eqnarray*} 
where $\bar{\alpha}$ stands for the variables $\rho$, $z$, $\tau$, and
the indices $a$, $b$, $c$ run over the values 1, 2, 3
(as always, there is the summation convention of repeated indices).

The transformations of the Higgs field
coefficient functions $H_j$ ($j=0,\ldots,3$) are the following :
\beq
 \left( \begin{array}{c} H_0 \\ H_1 \\ H_2 \\ H_3 \end{array} \right) \to
R_{\rm H} \left( \begin{array}{c} H_0 \\ H_1 \\ H_2 \\ H_3 \end{array} \right),
\label{htransf}
\eeq
with the $SO(4)$ matrix
\[
    R_{\rm H} = \cos\left({\frac{\Omega}{2}}\right) \id_4 +
    \sin\left({\frac{\Omega}{2}}\right)
    \frac{\vec{\omega} \cdot \vec{S}}{\Omega} \, ,
\]
defined in terms of the 4-dimensional identity matrix $\id_4$ and
\[
S_1 \equiv
\left( \begin{array}{cc} -i\sigma_2 & 0 \\ 0 & -i\sigma_2 \end{array} \right)
\qquad
S_2 \equiv
\left( \begin{array}{cc} 0 & -\sigma_3 \\ \sigma_3 & 0 \end{array} \right)
\qquad
S_3 \equiv
\left( \begin{array}{cc} 0 & -\sigma_1 \\ \sigma_1 & 0 \end{array} \right).
\]
 
\renewcommand{\theequation}{C.\arabic{equation}}
\setcounter{equation}{0}
\section*{Appendix C}

Here, we give the ansatz actiondensities $a_{\rm WKIN}$, $a_{\rm HKIN}$,
$a_{\rm HPOT}$ and Pontryagin density $q_{\rm P}$ in a compact,
transparent notation.

First, define the 3-dimensional coordinates 
\beq
(y_1, y_2, y_3) \equiv (\rho, z, \tau) \, .
\label{ycoordinates}
\eeq
\noindent Second, introduce the following
``isovectors'' given in terms of the coefficient functions $C_{i}$ of the
ansatz (\ref{eq:axial}) :
\begin{equation}
\vec{k}_1 \equiv \left( \begin{array}{c} C_1 \\ C_2 \\ C_6 \end{array} \right)
\qquad
\vec{k}_2 \equiv \left( \begin{array}{c} C_7 \\ - C_8 \\ C_9 \end{array} \right)
\qquad
\vec{k}_3 \equiv \left( \begin{array}{c} C_{10} \\ - C_{11} \\ C_{12} \end{array}\right)
\qquad
\vec{k}_4 \equiv \left( \begin{array}{c} C_3 \\ C_4 \\ 1-C_5 \end{array} \right)\, ,
\label{kvectors}
\end{equation}
which are motivated by the transformation properties (B.1) 
found in Appendix B.
\noindent Third, define the ``field strengths'' and ``covariant derivatives'' as
\begin{eqnarray}
\vec{K}_{\alpha \beta} &\equiv& 
\frac{\D}{\D y_{\alpha}} \left( \frac{\vec{k}_{\beta }} {y_{\beta}}  \right) -
\frac{\D}{\D y_{\beta}}  \left( \frac{\vec{k}_{\alpha}} {y_{\alpha}} \right) +
\frac{\vec{k}_{\alpha}}{y_{\alpha}} \times
\frac{\vec{k}_{\beta}} {y_{\beta}} \, ,
\\
D_{\alpha} \vec{k}_4   &\equiv&
\frac{\D}{\D y_{\alpha}} \vec{k}_4 +
\frac{\vec{k}_{\alpha}}{y_{\alpha}}  \times \vec{k}_4 \; ,
\end{eqnarray}
where the indices $\alpha$, $\beta$ take the values 1, 2, 3.
With these definitions the Yang-Mills actiondensity (\ref{eq:awkin}) and
Pontryagin density (\ref{eq:qp}) become simply
\begin{eqnarray}
a_{\mbox {\tiny WKIN}} &=& \frac{1}{2 g^2}          \left\{\;
\frac{1}{2} \sum_{\alpha, \beta = 1}^{3}
\left| \vec{K}_{\alpha \beta} \right|^2  + \, \sum_{\alpha = 1}^{3}
\left|\, \frac{1}{\rho}\,D_{\alpha} \, \vec{k}_4 \,\right|^2
                                                 \; \right\} \\
q_{\mbox {\tiny P}} &=& \frac{1}{2 \rho} \; \sum_{\alpha, \beta, \gamma=1}^{3}
\; \epsilon_{\alpha \beta \gamma}\;
\frac{\D}{\D y_{\gamma}} \left(\;  \vec{k}_{4} \cdot \vec{K}_{\alpha \beta} \; \right) \, . 
\end{eqnarray}

For the Higgs actiondensity it turns out to be useful to introduce
``4-vectors'', together with an implicit euclidean metric of positive signature.
First, define the following 4-vector  in terms of the coefficient
functions $H_{j}$ of the
ansatz (\ref{eq:axial}) :
\begin{equation}
H^{\mu} \equiv \left( \begin{array}{c} H_0\\H_1\\H_2\\H_3 \end{array}\right)\, .
\label{H4vector}
\end{equation}
Second, introduce the ``covariant derivatives''
\beq
{\cal D}_{\alpha} \equiv \frac{\D}{\D y_{\alpha}} \, \id_4 
+\, \frac{\vec{k}_{\alpha}}{y_{\alpha}} \cdot \frac{\vec{S}}{2} \;\; ,
\eeq
with $\alpha= 1,\,2,\, 3,$ and the $4 \times 4$ matrices
\beq
S_1 \equiv
\left( \begin{array}{cc} -i\sigma_2 & 0 \\ 0 & -i\sigma_2 \end{array} \right)
\qquad
S_2 \equiv
\left( \begin{array}{cc} 0 & -\sigma_3 \\ \sigma_3 & 0 \end{array} \right)
\qquad
S_3 \equiv
\left( \begin{array}{cc} 0 & -\sigma_1 \\ \sigma_1 & 0 \end{array} \right),
\eeq
which appeared already in the transformations (B.2) 
of Appendix B.  Third, define the matrix
\beq
K_4 \equiv  \vec{k}_4 \cdot \vec{S} + T_3 \, ,
\eeq
where
\beq
T_3 \equiv
\left( \begin{array}{cc} 0 & i\sigma_2 \\ i\sigma_2 & 0 \end{array} \right)
\eeq
commutes with all $S_a$.
With these definitions the Higgs actiondensities (\ref{eq:ahkin}) and
(\ref{eq:ahpot})  become
\begin{eqnarray}
a_{\mbox {\tiny HKIN}} &=& \frac{v^2}{2}\,                  \left\{\;
\sum_{\alpha=1}^{3} \; 
\left( \, {\cal D}_{\alpha}^{\, \mu \nu} \, H^{\nu}\,\right)^2 +
\left( \; \frac{1}{2 \rho} \; K_4^{\mu\nu} H^{\nu} \right)^2       \;\right\} \\
a_{\mbox {\tiny HPOT}} &=& \lambda\, \frac{v^4}{4}
\left(\, H^{\mu} H^{\mu}  - 1 \, \right)^2 \, ,
\end{eqnarray}
where the last term in $a_{\mbox {\tiny HKIN}}$ mixes the two types of ``scalars''
($\vec{k}_{4}$ and $H^{\mu}$) of the effective $SO(3)$ \YMHth ~found.

\end{appendix}

\end{document}